\documentclass[superscriptaddress,secnumarabic,amssymb,amsmath,nobibnotes,aps,prd,showkeys,showpacs,nofootinbib,preprint]{revtex4}

\usepackage{graphicx}
\usepackage{float}
\usepackage{bm}
\usepackage{amsmath}
\usepackage{amsfonts}
\usepackage{amssymb}
\usepackage{epstopdf}
\usepackage{natbib}%
\newcommand{\bee}{\begin{equation}}
\newcommand{\eee}{\end{equation}}
\newcommand{\eaa}{\end{eqnarray}}
\newcommand{\baa}{\begin{eqnarray}}
\def\ni{\noindent}

\usepackage{color}

\begin{document}

\title{Black holes quasinormal modes, Loop Quantum Gravity Immirzi parameter and nonextensive statistics}

\author{Everton M. C. Abreu}\email{evertonabreu@ufrrj.br}
\affiliation{Departamento de F\'{i}sica, Universidade Federal Rural do Rio de Janeiro, 23890-971, Serop\'edica, RJ, Brazil}
\affiliation{Departamento de F\'{i}sica, Universidade Federal de Juiz de Fora, 36036-330, Juiz de Fora, MG, Brazil}
\affiliation{Programa de P\'os-Gradua\c{c}\~ao Interdisciplinar em F\'isica Aplicada, Instituto de F\'{i}sica, Universidade Federal do Rio de Janeiro, 21941-972, Rio de Janeiro, RJ, Brazil}
\author{Jorge Ananias Neto}\email{jorge@fisica.ufjf.br}
\affiliation{Departamento de F\'{i}sica, Universidade Federal de Juiz de Fora, 36036-330, Juiz de Fora, MG, Brazil}
\author{Ed\'esio M. Barboza Jr.}\email{edesiobarboza@uern.br}
\affiliation{Departamento de F\'isica, Universidade do Estado do Rio Grande do Norte, 59610-210, Mossor\'o-RN, Brazil}
\author{Br\'aulio B. Soares}\email{brauliosoares@uern.br}
\affiliation{Departamento de Ci\^encia e Tecnologia, Universidade do Estado do Rio Grande do Norte, Natal, RN, Brazil}

%%%%%%%%%%%%%%%%%%%%%%%%%%%%%%%%%%%%%%%%%%%%%%%%%%%%%%%%%%%%%%%%%%%%%%%%%%%%%%%%%%%%%%%%%%%%

%%%%%%%%%%%%%%%%%%%%%%%%%%%%%%%%%%%%%%%%%%%%%%%%%%%%%%%%%%%%%%%%%%%%%%%%%%%%%%%%%%%%%%%%%%
\begin{abstract}
\noindent It is argued that, using the black hole area entropy law together with the Boltzmann-Gibbs statistical mechanics and the quasinormal modes of the black holes, it is possible to determine univocally the lowest possible value for the spin $j$ in the context of the Loop Quantum Gravity theory which is $j_{min}=1$. Consequently, the value of Immirzi parameter is given by $\gamma = \ln 3/(2\pi\sqrt{2})$.  In this paper, we have shown that if we use Tsallis microcanonical entropy rather than Boltzmann-Gibbs framework then the minimum value of the label $j$ depends on the  nonextensive  $q$-parameter and may have values other than $j_{min}=1$. 
\end{abstract}
%%%%%%%%%%%%%%%%%%%%%%%%%%%%%%%%%%%%%%%%%%%%%%%%%%%%%%%%%%%%%%%%%%%%%%%%%%%%%%%%%%%%%%%%%%%%
\date{\today}

\maketitle
%%%%%%%%%%%%%%%%%%%%%%%%%%%%%%%%%%%%%%%%%%%%%%%%%%%%%%%%%%%%%%%%%%%%%%%%%%%%%%%%%%%%%%%%%%%%%%
 Loop Quantum Gravity (LQG)\cite{lqgr} is a theory of quantum gravity that proposes to unify quantum field theory and general relativity. The main attractive point in LQG is, in principle, the possibility to describe the quantum spacetime in a nonperturbative background-independent form.
The Hilbert space of LQG is formed by spin networks which are graphs with edges that carry labels such as $j=0, 1/2, 1, 3/2,\ldots\,$. In LQG, the area of a given region of space has a discrete spectrum in such a way that, if a surface is intersected or punctured by the spin network edge that carries the label $j$, then the surface carries an area element written as \cite{rs,al,ms}
 \begin{eqnarray}
\label{a1}
a(j)=8 \pi l_p^2\, \gamma\, \sqrt{j(j+1)}\,\,\,,
\end{eqnarray}

\ni where $l_p$ is the Planck length and $\gamma$ is the so-called Immirzi parameter \cite{immi}.    Eq. (\ref{a1}) is an relevant prediction of LQG. However, this important relation is weakened by the fact that the Immirzi parameter is, in principle, undetermined. By definition, the Immirzi parameter carries the measure of the size of the quantum of area in Planck's units.
One way to compute the Immirzi parameter, by solving the problem mentioned above, can be carried out with the help of quasinormal modes in the black holes theory. Quasinormal modes are damped oscillations that appear in the perturbation equations of the Schwarzchild geometry. These solutions were initially found by Regge and Wheeler \cite{rw}. This procedure, as we will see, connects the relation between area and mass of a Schwarzschild black hole to the area produced by the spin network in the context of LQG \cite{od}.
On the other hand, the interested reader can notice that an interesting work \cite{wr}, using a conformal gauge structure in a novel generalized Holst action, obtains non-fixed values of the Immirzi parameter, i.e., the Immirzi parameter may be not beset by ambiguities in this approach.   

In this paper, following the work of Dreyer \cite{od},  we will use the quasinormal modes to obtain a new equation for  the minimum value of the spin $j$ that appears in Eq.(\ref{a1}) as a consequence of a nongaussian statistics, namely, Tsallis' statistics.    It is important to mention that in Ref. \cite{od} the author has considered Boltzmann-Gibbs (BG) statistics. 

Tsallis' formalism \cite{tsallis}, which is an extension of BG statistical theory, defines a nonextensive (NE), i.e., a nonadditive entropy such as
\begin{eqnarray}
\label{nes}
S_q =  k_B \, \frac{1 - \sum_{i=1}^W p_i^q}{q-1}\;\;\;\;\;\;\qquad \Big(\,\sum_{i=1}^W p_i = 1\,\Big)\,\,,
\end{eqnarray}

\ni where $p_i$ is the probability of a system to exist within a microstate, $W$ is the total number of configurations (microstates) and 
$q$, known in the current literature as the Tsallis parameter or NE  parameter, is a real (or not \cite{abreu}) parameter which measures the degree of nonextensivity. 
The definition of entropy in Tsallis statistics carries the standard properties of positivity, equiprobability, concavity and irreversibility. This approach has been successfully used in many different physical systems. For instance, we can mention the Levy-type anomalous diffusion \cite{levy}, turbulence in a pure-electron plasma \cite{turb} and gravitational systems \cite{sys,sa,eu,maji,mora}.
It is noteworthy to affirm that Tsallis thermostatistics formalism has the BG statistics as a particular case in the limit $ q \rightarrow 1$ where the standard additivity of entropy can be recovered. 

In the microcanonical ensemble, where all the states have the same probability, Tsallis entropy reduces to \cite{sys}
\begin{eqnarray}
\label{micro}
S_q=k_B\, \frac{W^{1-q}-1}{1-q},
\end{eqnarray}
where in the limit $q \rightarrow 1$ we recover the usual Boltzmann entropy formula, $S=k_B\, \ln {W}$.

For a large imaginary part of the quasinormal frequency, which will be denoted as $\omega$, Nollert \cite{no} has obtained the following limiting of the quasinormal mode frequencies 
\begin{eqnarray}
\label{nmw}
M \omega_n = 0.04371235 + \frac{i}{4} \left( n + \frac{1}{2} \right)\,,
\end{eqnarray}

\ni where  $M$ is the mass of black hole and $n$ is a non-negative integer. Here we are using the  gravitational units $G=c=1$. The asymptotic behavior Eq.(\ref{nmw}) was later verified by
Anderson \cite{and} using for this an independent analysis. An important observation was made by Hod \cite{hod} where the real numeric constant of Eq.(\ref{nmw}) is equal to
\begin{eqnarray}
\label{nc}
{\it Re}[M\omega_n]=\frac{\ln 3}{8 \pi}\,.
\end{eqnarray}

\ni Denoting the real part of the quasinormal modes as $w_n$ then from Eqs. (\ref{nmw}) and (\ref{nc}) we have
\begin{eqnarray}
\label{omega}
w_n = \frac{\ln 3}{8 \pi M}.
\end{eqnarray}

\ni Moreover, based on the considerations made by Hod \cite{hod} we can assume that the quantum of energy is
\begin{eqnarray}
\label{de}
\Delta M = E = \hbar w_n \,.
\end{eqnarray}

\ni From Eqs. (\ref{omega}) and (\ref{de}) we have that
\begin{eqnarray}
\label{dm}
\Delta M = \frac{\hbar \ln 3}{8 \pi M} \,.
\end{eqnarray}

\ni Using the usual relation for a Schwarzschild black hole 
\begin{eqnarray}
\label{ab}
A = 16 \pi M^2 \,,
\end{eqnarray}

\ni the mass change of Eq.(\ref{dm}) translates into the area change given by
\begin{eqnarray}
\label{da1}
\Delta A = 4 \ln 3 \;  l_p^2 \,,
\end{eqnarray}

\ni where we have used that $\hbar = l_p^2$ in gravitational units. On the other hand, by using the area result of LQG, Eq.(\ref{a1}),  we have

\begin{eqnarray}
\label{da}
\Delta A = a(j_{min}) = 8 \pi l_p^2 \gamma \sqrt{j_{min}(j_{min}+1)} \,.
\end{eqnarray}

\ni Comparing Eq.(\ref{da}) with (\ref{da1}) we can obtain an expression for the
Immirzi parameter $\gamma$ given by
\begin{eqnarray}
\label{gama}
\gamma=\frac{\ln 3}{2 \pi \sqrt{j_{min} (j_{min}+1)}} \,.
\end{eqnarray}

In order to derive the dependence of the minimum spin $j$, we will consider the Tsallis statistics. The number of configurations (microstates) in a punctured surface is given by
\begin{eqnarray}
\label{nconf}
W=\prod_{n=1}^{N} (2 j_n+1)\,\,,
\end{eqnarray}

\ni where the term $(2 j_n+1)$ in Eq. (\ref{nconf}) is the multiplicity of the state $j$. Statistically, the most important configurations that contribute into Eq. (\ref{nconf}) come from   $j_n=j_{min}$, where $j_{min}$ is the minimum label.   Then, from Eq. (\ref{nconf}),  we have that
\begin{eqnarray}
\label{micron}
W=(2 j_{min}+1)^N\,\,.
\end{eqnarray}

\ni From Eqs. (\ref{a1}) and (\ref{da})  the number $N$ of punctures in a surface of area $A$ is given by

\begin{eqnarray}
\label{np}
N=\frac{A}{a(j_{min})}=\frac{A}{\Delta A}\,\,.
\end{eqnarray}

\ni From (\ref{da1}) we can write Eq.(\ref{np}) as
\begin{eqnarray}
\label{npm}
N=\frac{A}{4 l_p^2 \ln 3}\,.
\end{eqnarray}

\ni By replacing Eqs. (\ref{micron}) and (\ref{npm}) into (\ref{micro}) we obtain
\begin{eqnarray}
\label{microjmin}
S_q=k_B\, \frac{(2j_{min}+1)^{(1-q)\frac{A}{4 l_p^2 \ln 3}}-1}{1-q}.
\end{eqnarray}
In order to solve this equation for $j_{min}$, we make $S_q$ equal to the black hole entropy, i. e., $S_q=k_BA/(4l_p^2)$ so that

\begin{eqnarray}
\label{jmin}
j_{min} = \frac{\Big[ 1+(1-q) \frac{ A}{4 l_p^2} \Big]^{\frac{ \ln 3}{(1-q) \frac{A}{4l_p^2}}}- 1}{2} \,.
\end{eqnarray}

\ni It is possible to show that $\lim_{q\to1}j_{min} = 1$ (see also Figure 1)  which is the result obtained by Dreyer \cite{od}. However, the degree of freedom introduced by the parameter $q$ allows $j_{min}$  take different values. Particularly, $j_{min}=1/2$ as
\begin{equation}
\label{jmin.5}
1-q\approx\ln 3\Big[\frac{5-18\ln(3/2)}{4\ln 3-9\ln 2}\Big]\frac{4l_p^2}{A}\approx1.37\frac{4l_p^2}{A}.
\end{equation}
Since the entropy concavity requires that $q>0$ \cite{tsallis}, the value $j_{min}=1/2$ it is allowed only to black holes of area $A\gtrsim5.48l_p^2$. Also, $j_{min}\to\infty$ when $(1-q)\to -4l_p^2/A$. Thus, physically acceptable values of $q$ lies in the range
\begin{equation}
\label{qdomain}
  -4l_p^2/A<(1-q)\lesssim1.37\frac{4l_p^2}{A},\quad A\gtrsim5.48l_p^2.
\end{equation}  
Since a micro black hole has an area of $16\pi l_p^2$ the condition (\ref{qdomain}) is satisfied by any physically realizable black hole. In Figure 1 we have plotted the minimum value for the spin $j$, Eq.(\ref{jmin}), as a function of $\kappa\equiv 1-q$ for a micro black hole, $A=16\pi l_p^2$, and for $A=32\pi l_p^2$. The circle ($16\pi l_p^2$) and square ($32\pi l_p^2$) points  show the $\kappa$ parameters corresponding to $j_{min}$ values usually allowed which are positive half-integers, i.e., $j_{min}=1/2,1,3/2,2, 5/2,...$. For the case $A=16\pi l_p^2$, the physical constraint (\ref{qdomain}) implies that $-0.08<\kappa\lesssim0.11$. This range is shortened by a factor of two for $A=32\pi l_p^2$. In fact, the $\kappa$ interval can be made as small as you want by increasing sufficiently the black hole area. This means that, for any realizable black hole, a small deviation of the standard BG statistics in the Tsallis formalism can accommodate easily the usual case $j_{min}=1/2$. Also, we can see that values of $q>1(\kappa<0)$ lead to values of $j_{min}>1$.

\begin{figure}[H]
	\centering
	\includegraphics[width=14.0cm]{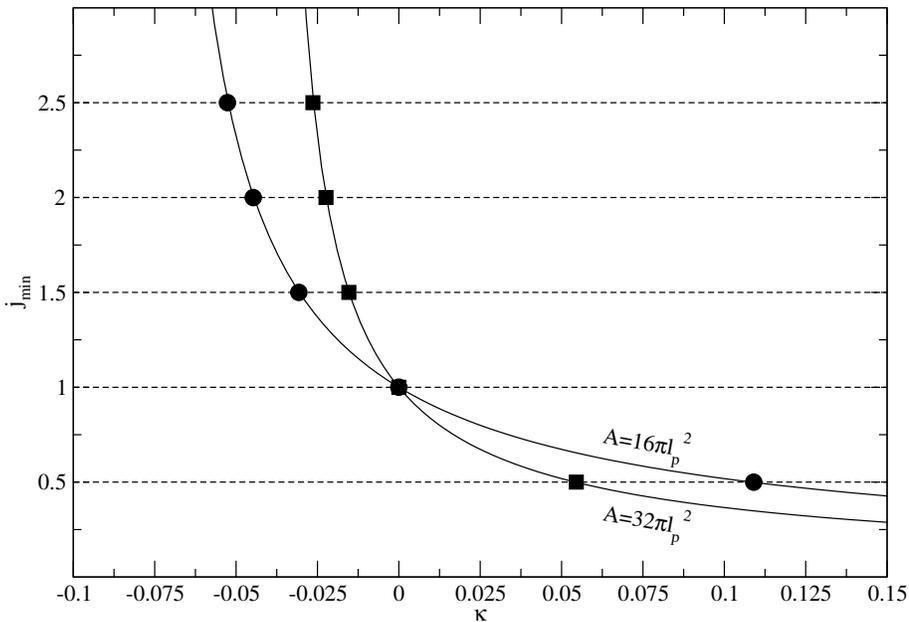}
	\caption{Values of $j_{min}$ as a function of $\kappa\equiv 1-q$ for micro black holes of areas $A=16\pi l_p^2$ and $32\pi l_p^2$.}
	\label{min}
\end{figure}

%\ni From Fig. 1 we can observe that for  the value of the minimum spin $j_{min}$ decreases when the value of $\kappa$ increases.
%Also from Fig. 1 we can observe that for $\kappa < 0.25$ which corresponds $q<1$  we have $j_{min}=\frac{1}{2}$ which is the usual value adopted for $j_{min}$. 

An expression for the Immirzi parameter as a function of $q$ is obtained from Eq.(\ref{gama}) as
\begin{eqnarray}
\label{gamaj}
\gamma= \frac{\ln 3}{\pi \sqrt{ \Big[1+ (1-q)\frac{ A}{4 l_p^2}\Big]^{\frac{2 \ln 3}{(1-q) \frac{A}{4 l_p^2}}}-1} } \,,
\end{eqnarray}

\ni which recovers the result obtained by Dreyer \cite{od}, $\gamma= \frac{\ln 3}{2 \pi \sqrt{2}}$, 
in the limit $q \rightarrow 1$. 

%\ni which is equal to the result obtained by Dreyer \cite{od}.
Figure 2 shows the Immirzi parameter as a function of $\kappa$ for  $A=16\pi l_p^2$ and $A=32\pi l_p^2$.

\begin{figure}[H]
	\centering
	\includegraphics[width=14.0cm]{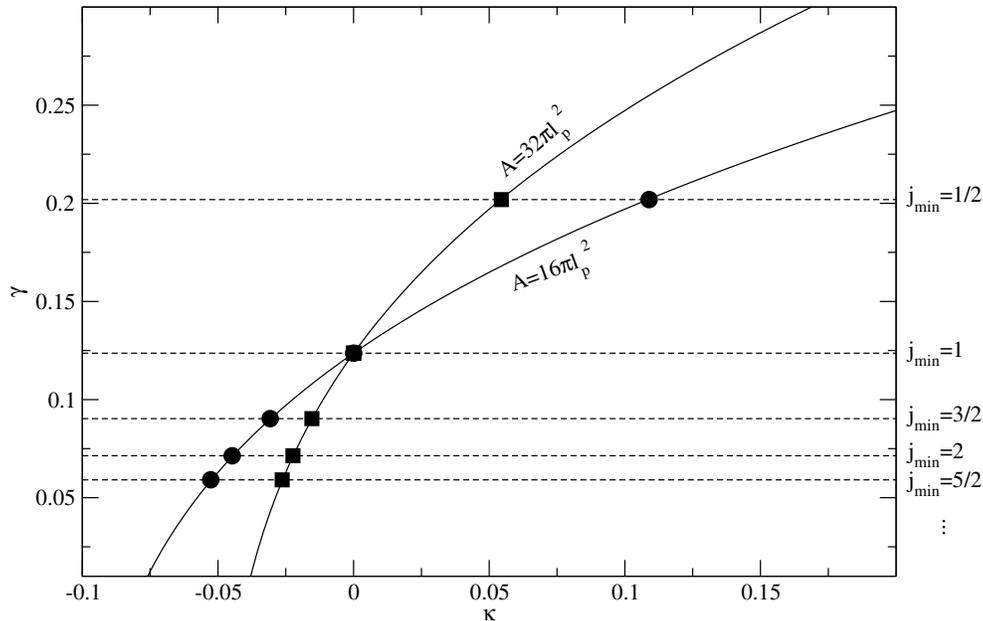}
	\caption{The Immirzi parameter $\gamma$ as a function of $\kappa\equiv 1-q$ for micro black holes of areas $A=16\pi l_p^2$ and $32\pi l_p^2$. The dashed horizontal lines show the values of $\gamma$ corresponding to the first five possible values of $j_{min}$.}
	\label{gamag}
\end{figure}

\ni As we can see that  the value of the Immirzi parameter $\gamma=\frac{\ln 3}{\pi\sqrt{3}}$, corresponding to $j_{min}=1/2$, can be attained for small deviations of $q=1$ for any physically realizable black hole. Here it is important to mention that, analogously to Dreyer's model \cite{od}, when we use Eq. (12), the standard value $\gamma=0.274$ \cite{alepz} does not take the minimum spin $j$ to be semi-integer or integer, i.e., $j\approx 0.331$, and,  consequently, this usual value of the Immirzi parameter can not be reproduced by our model.  However, our main result, Eq. (21), remains  correct because we have imposed  that the Tsallis entropy, Eq. (17), is equal to the black holes entropy. The strategy of reproducing black holes entropy, which results in the obtainment of the Immirzi parameter,  is a standard procedure in LQG.

To conclude, in this work we have investigated the behavior of the lowest possible spin, $j_{min}$, in the framework of LQG when we consider Tsallis entropy, Eq.(\ref{micro}), instead of BG entropy. Our result, which is inside Eq.(\ref{jmin}), shows that the minimum spin number depends on the ratio $A/4l_p^2$ and the nonextensive parameter $q$. In the limit $q \rightarrow 1$, where BG framework must be recovered, we have reobtained the result $j_{min}=1$.
It is important to mention that Dreyer \cite{od}, considering only the BG entropy, has obtained that the lowest possible spin is $j_{min}=1$. Therefore he has concluded that the gauge group of the spin networks in the context of LQG is $SO(3)$ and not $SU(2)$. However, if we consider Tsallis' entropy, we can show that both $SU(2)$, which it has normally been adopted, as well $SO(3)$, which it has been claimed by Dreyer, are possible gauge groups for LQG. Also, we show values of the entropic parameter $q>1$ lead to $j_{min}>1$. 
Then, Tsallis' statistics can certainly generalize the value of the lowest spin possible $j_{min}$. 

The Immirzi parameter is also obtained with the aid of Eqs. (\ref{gama}) and (\ref{jmin}) and the result is Eq.(\ref{gamaj}).  As expected, in the limit $q \rightarrow 1$, where BG must be recovered, we have obtained $\gamma=\frac{\ln 3}{2\pi\sqrt{2}}$. Therefore, our result indicates that if we consider an important nongaussian statistics, which is Tsallis' entropy, instead of BG statistics, then the minimum  possible value of the spin networks and the Immirzi parameter have a plethora of values possible.

\section*{Acknowledgments}

\ni This research was financially supported in part by the Coordena\c{c}\~ao de Aperfei\c{c}oamento de Pessoal de N\'ivel Superior - Brasil (CAPES) - Finance Code 001. The authors thank CNPq (Conselho Nacional de Desenvolvimento Cient\' ifico e Tecnol\'ogico), Brazilian scientific support federal agency, for partial financial support, Grants numbers  406894/2018-3 and 302155/2015-5 (E.M.C.A.) and 303140/2017-8 (J.A.N.).

\end{document}